\documentclass{article}

\PassOptionsToPackage{numbers, compress}{natbib}

\usepackage[final]{neurips_2023_ml4ps}
\usepackage{float}

\usepackage{graphicx}
\usepackage[utf8]{inputenc} 
\usepackage[T1]{fontenc}    
\usepackage{hyperref}       
\usepackage{url}            
\usepackage{booktabs}       
\usepackage{amsfonts}       
\usepackage{nicefrac}       
\usepackage{microtype}      
\usepackage{xcolor}         

\title{Reconstruction of Fields from Sparse Sensing: Differentiable Sensor Placement Enhances Generalization}

\author{%
  Agnese Marcato \\
  Earth and Enviromental Sciences\\
  Los Alamos National Laboratory,\\
  Politecnico di Torino\\
  \texttt{agnese.marcato@polito.it} \\
  \And
  Daniel O'Malley \\
  Earth and Enviromental Sciences\\
  Los Alamos National Laboratory\\
  \texttt{omalled@lanl.gov} \\
  \AND
  Hari Viswanathan \\
  Earth and Enviromental Sciences\\
  Los Alamos National Laboratory\\
  \texttt{viswana@lanl.gov} \\
  \And
  Eric Guiltinan \\
  Earth and Enviromental Sciences\\
  Los Alamos National Laboratory\\
  \texttt{eric.guiltinan@lanl.gov} \\
  \And
  Javier E. Santos \\
  Earth and Enviromental Sciences\\
  Los Alamos National Laboratory\\
  \texttt{jesantos@lanl.gov} \\
}
\begin{document}

\maketitle

\begin{abstract}
  Recreating complex, high-dimensional global fields from limited data points is a grand challenge across various scientific and industrial domains. 
  Given the prohibitive costs of specialized sensors and the frequent inaccessibility of certain regions of the domain, achieving full field coverage is typically not feasible. Therefore, the development of algorithms that intelligently improve sensor placement is of significant value. In this study, we introduce a general approach that employs differentiable programming to exploit sensor placement within the training of a neural network model in order to improve field reconstruction. 
  We evaluated our method using two distinct datasets; the results show that our approach improved test scores. Ultimately, our method of differentiable placement strategies has the potential to significantly increase data collection efficiency, enable more thorough area coverage, and reduce redundancy in sensor deployment.
\end{abstract}

\section{Introduction}

The challenge of reconstructing spatial fields that change over time from limited sensor data has been a focal point for many research studies~\cite{wang2013confident,jiang2022online, manohar2018data}. Various machine learning methods have been used in attempts to address this complex issue, including several types of neural networks. For instance, shallow neural networks~\cite{erichson2020shallow}, as well as convolutional neural networks~\cite{fukami2021global}, have been employed. More recently, attention-based neural networks~\cite{santos2023development} have been investigated. Despite their different architectures, all of these methods share the common requirement that the user needs to manually determine the sensor positions. This requirement remains a limiting factor in the ongoing quest for efficient learning and accurate field reconstruction.

The strategic placement of sensors is perhaps the most critical aspect of sparse sensing. It deeply influences the performance and efficiency of data collection systems and, consequently, the methods used to reconstruct data. 
Considering the non-linear nature of physical fields and training algorithms, this optimization problem proves to be quite challenging, especially when the objective is to obtain a flexible model for different applications.

This study aims to present a method that enables a model to reconstruct a field by optimizing the sensors positions via backpropagation, thereby facilitating the model's exploration of the spatial domain and enhancing sensor positioning effectively. However, to accomplish this, the model must be allowed to index sensor values. Indexing, a basic operation in computer science, naturally incorporates discrete decisions, such as selecting a particular element from an array based on an index. This operation is non-differentiable since it doesn't yield a continuous output that can be smoothly adjusted with respect to its input, which is a requirement for the application of gradient-based optimization methods.

\begin{figure*}[h!]
\centering
\includegraphics[width=1\textwidth]{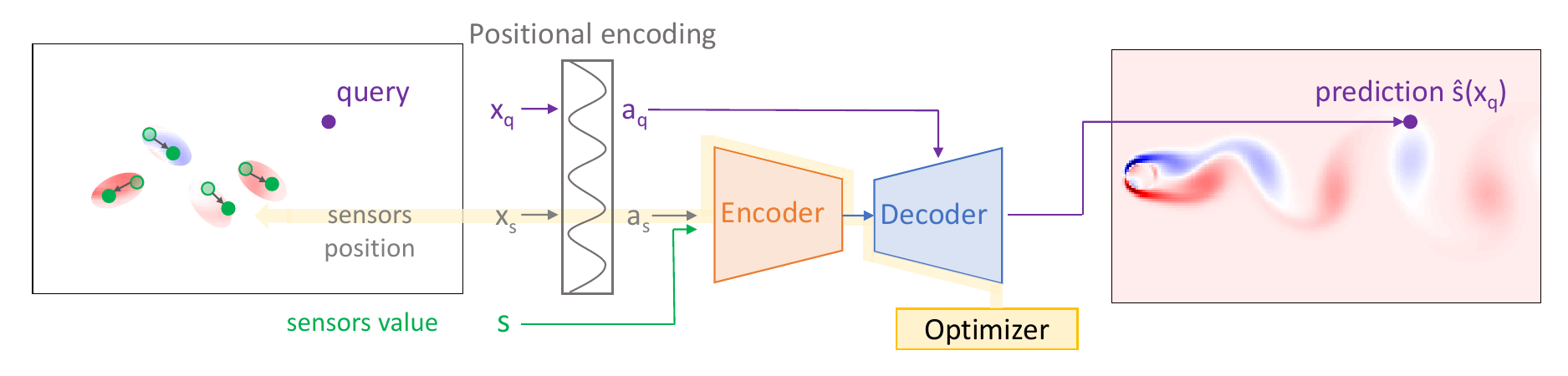}
    \caption{A schematic representation of the proposed end-to-end differentiable framework for sensor placement improvement. This workflow diagram showcases the key stages, from sparse representation to the final state, highlighting sensor movement during training.}
    \label{fig:workflow}
\end{figure*}

To overcome this challenge, we introduce a fully-differentiable model that enables gradient-based optimization methods to traverse an n-dimensional field using a set of discrete points. Our model utilizes interpolation and sine-cosine positional encodings to allow the sensors to move through the permissible space. 
We showcased its effectiveness by training an attention-based neural network, which achieved top-tier performance on two separate datasets.
The notion of "permissible space" is also explored: beyond the usual cartesian domain boundaries, arbitrarily defined internal boundaries are also expressed, representing real world unpredicted domain inaccessibility to sensor placement.
To our knowledge, this represents the first fully end-to-end differentiable workflow for exploiting the sensor placement process within a neural network model. A diagram illustrating our workflow is shown in Fig.~\ref{fig:workflow}.

\subsection{Methods}
While our workflow is general and could be applied with any neural network architecture, we chose to implement it in an attention based neural network \cite{ jaegle2021perceiverio, vaswani2017attention}. 
The main advantages of using an attention-based neural network include its inherent capability to grasp long-range dependencies between distant elements, the use of dynamic receptive fields that adapt based on input data \cite{jaegle2021perceiver}, and a design that does not rely as much on spatial structures~\cite{bello2019attention}. 

The attention based neural network we implemented in our optimization framework is the Senseiver~\cite{santos2023development} that has proven to be highly effective for the task of sparse reconstruction.
The Senseiver is trained to learn compact representations of system states from a small number of sensor observations at a given time, that is the input of the model. Then, the encoded representation is used to decode the state of the full system for new queries. The model inputs are a set of sensor observations at a given time. 

The attention based neural network model, as depicted in Fig. \ref{fig:workflow_opt}, comprises three primary elements: 1. A positional encoder which translates a spatial coordinate  into a set of spatial encodings facilitating the conversion of an n-dimensional spatial location into a vector. 2. An attention-driven encoder which processes the spatial encodings of sensor locations and their corresponding values, yielding a latent matrix. This matrix serves as a condensed representation of the system at a specific time. 3. An attention-focused decoder that predicts the field value at any given query position.

Initially, we determine the number of sensors whose locations we aim to optimize. These locations, along with the neural network's weights, are treated as trainable parameters within the model, allowing the optimizer, Adam \cite{kingma2014adam}, to adjust them throughout the training process. 
After every step of optimization, the locations of the sensors are adjusted. Consequently, it becomes necessary to recalculate both the positional encoding and the values of the property at these updated positions. This recalculation is integral to the continuation of the training process. To ensure seamless integration into the training workflow, these recalculations must be performed in a differentiable manner. The specifics of this differentiable approach are elaborated in the subsequent paragraph.

\subsection{Differentiable sensor placements}
A naive approach to get the property values for the new sensors position  would be to use the indexed discrete positions as trainable parameters, round them to the nearest integer and utilize these to index the field. However, this process, referred to as array indexing, is non-differentiable.

To overcome this limitation, and allow gradient flow, we implemented a differentiable interpolation method to determine the sensors values at precise locations. In particular, given the updated positions of the sensors at each step, the interpolation scheme is employed to compute the values of the property at those points, based on their neighboring data. These calculated sensor values then serve as inputs to our neural network. Since the model also relies on knowing the sensor positions, we converted their precise coordinates into a uniformly sized vector using Fourier positional encodings~\cite{vaswani2017attention}. 
To prevent sensors from entering prohibited zones within the domain, we introduced a bounce-back mechanism. Specifically, when a sensor's location comes within a one-pixel distance from the restricted area, its gradient direction is reversed, ensuring the sensor moves away from the boundary.
For a complete overview of the optimization workflow and the architecture details the reader can refer to the Appendix.

\section{Results}

The initial dataset we analyzed was derived from a simulation of two-dimensional unsteady flow navigating a cylindrical obstacle~\cite{colonius2008fast}. The resulting simulation manifests a phenomenon known as the von Kármán vortex street – an alternating pattern of left- and right-handed vortices in the flow field trailing the cylinder. The simulation, which solves the incompressible Navier-Stokes equation at a Reynolds number of 100, was conducted on a computational grid of 192x112 with 5,000 time frames, which roughly corresponds to four periods of vortex shedding. The simulation domain extends 10 cylinder diameters vertically and 15.7 cylinder diameters horizontally. The focus of data collection was the fluid's vorticity field, influenced by the presence of a solid obstruction. During the training phase, we randomly selected 50 frames, amounting to 1\% of the total data. The training is stopped when a plateau in the training loss is reached.

We trained the attention-based neural network optimizing the position of the sensors within the training (\textit{moving} strategy) and keeping their position fixed (\textit{fixed} strategy).We trained the system using 4, 8, and 16 sensors, and we evaluated their initialization across 4 distinct scenarios: literature locations~\cite{fukami2021global} (strategic points where the field presents high variation) and three random locations within the domain.

\begin{figure*}[h!]
\centering
\includegraphics[width=0.7\textwidth]{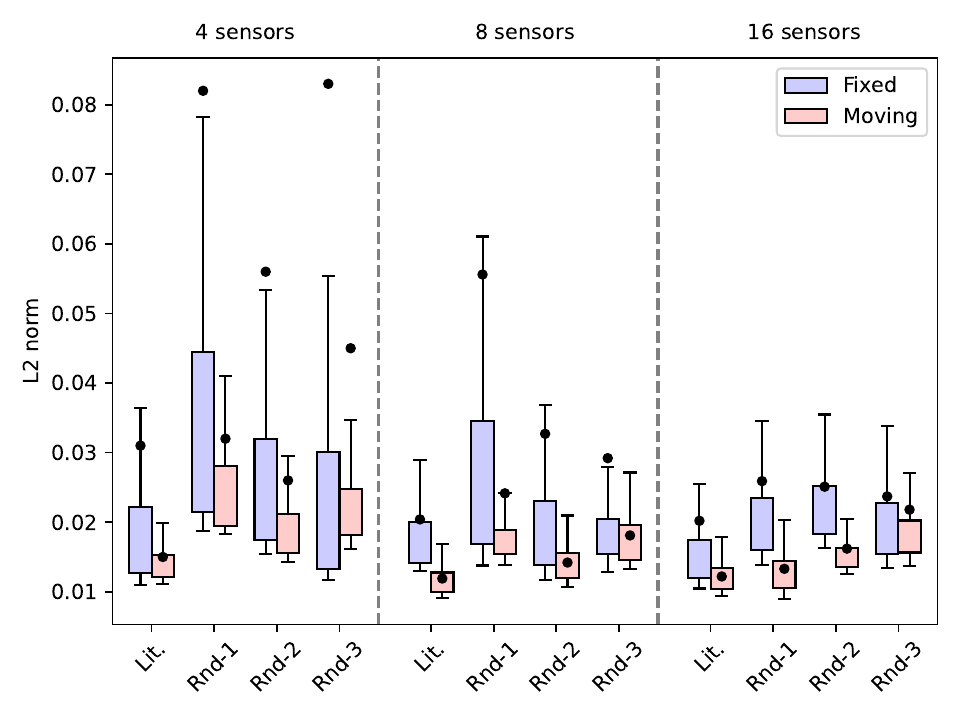}
    \caption{Error in the reconstruction of the vorticity field for the two sensor strategies, four different initialization of the sensors location and an increasing number of sensors (4,8,16). Box and whiskers plots, and arithmetic mean of the error (bullet). }
    \label{fig:cylinder_whisker}
\end{figure*}

The results are summarized in the plot of Fig. \ref{fig:cylinder_whisker}. 
The performance of the trained models was assessed using the remaining 4,950 frames from the dataset using the L$_2$ norm ($\frac{ \|  y-\hat{y} \|_{2}}  {\| y \|_{2}}$). As expected, increasing the number of sensors the reconstruction performance improves. The moving sensor strategy highly improves the field reconstruction capability of the network, in terms of mean and standard deviation of the error distribution, as seen in the comparison of the percentiles.

\begin{figure*}[h!]
\centering
\includegraphics[width=0.7\textwidth]{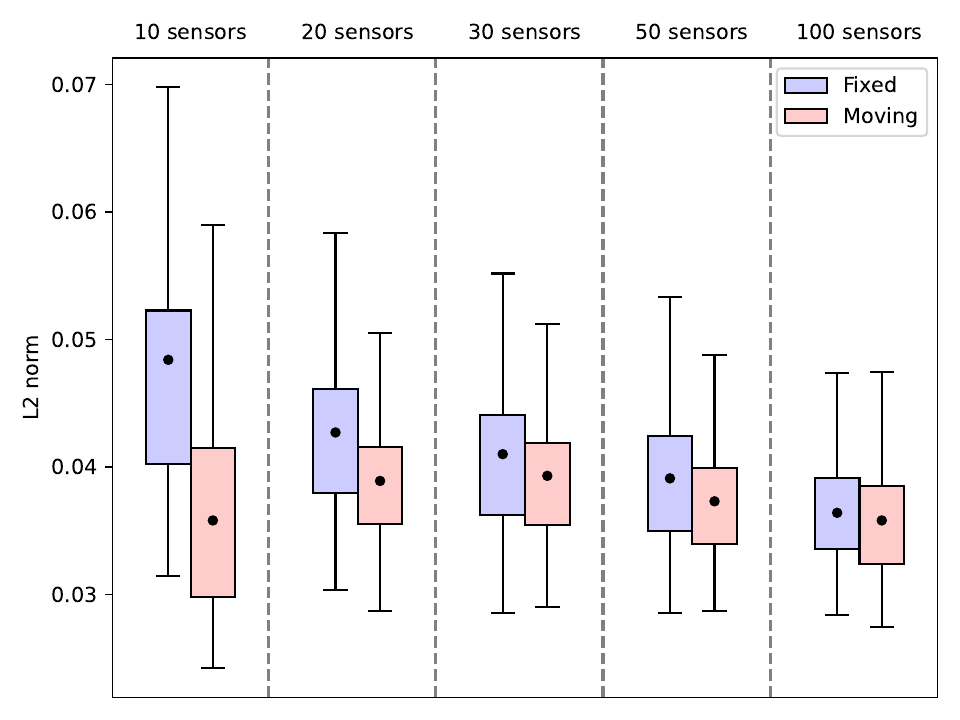}
    \caption{Error in the reconstruction of the NOAA temperature field for the two sensor strategies and an increasing number of sensors (10, 20, 30, 50, 100). Box and whiskers plots, and arithmetic mean of the error (bullet).}
    \label{fig:toset}
\end{figure*}

\begin{figure*}
    \centering
    \includegraphics[width=0.90\textwidth]{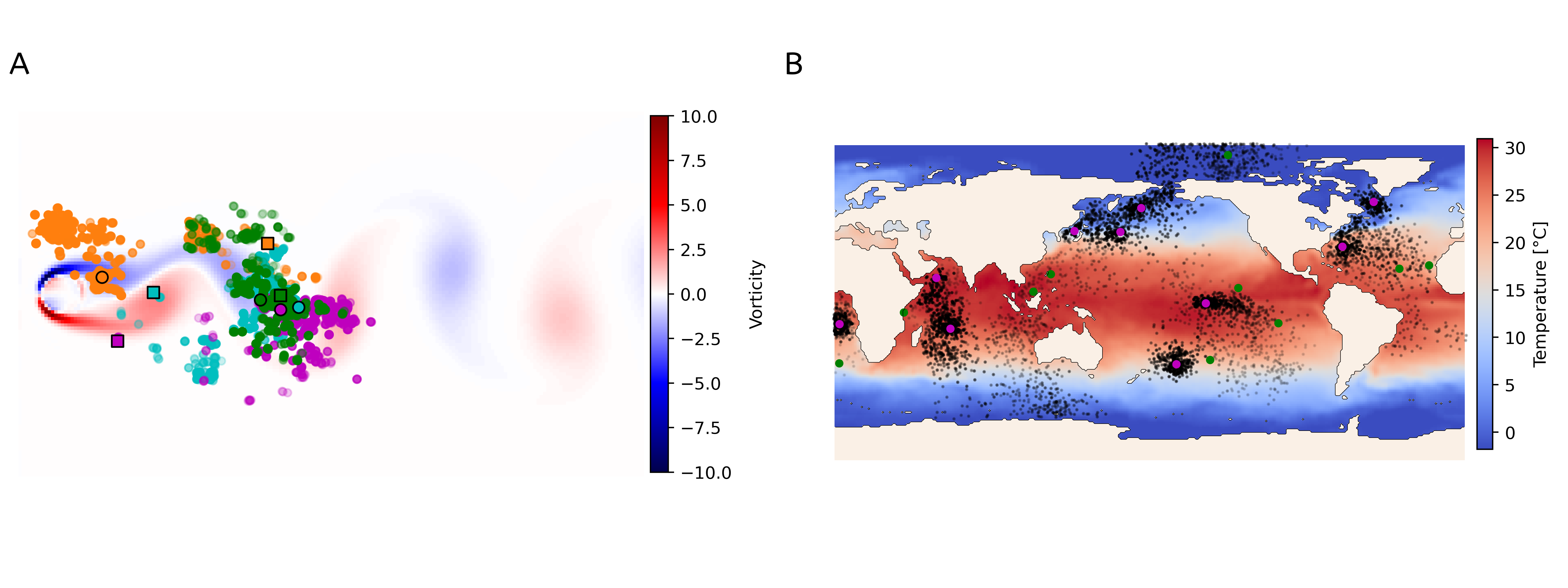}
    \caption{\textbf{A}. Initial positions (square) and final position post-training (circles) for the 4 sensors in the flow past a cylinder case. \textbf{B}. Initial sensor positions are indicated by green markers, with their post-training final positions denoted in purple. The paths traveled by the sensors during training are represented by the trails of black dots. It's noteworthy that, due to the momentum component in the Adam optimizer, sensors can travel inland.}
    \label{fig:seat_temp}
\end{figure*}

The second dataset that we considered is the NOAA sea surface temperature~\cite{NOAA}. This real-world dataset was collected from satellite and ship-based observations over time. The data comprise weekly observations of the sea surface temperature of the Earth at a spatial resolution of 360x180 (longitude and latitude, respectively) -- giving a resolution of one degree longitude and one degree latitude. For training, we use 1,040 snapshots spanning from the year 1981 to 2001, and then we tested on snapshots from 2001 to 2018. An increasing number of sensors has been studied, and the moving and fixed training strategy has been tested for random initialization of the locations. The results are analogous to the previous ones and are reported in Fig.\ref{fig:toset}.

For both scenarios, dynamically optimizing the sensor positions during the training process resulted in a decrease in the average test error, calculated across all frames, as well as in the other metrics. This significant reduction in error rates strongly indicates that our dynamic optimization approach considerably enhances the generalization capability of our model. A schematic of how the sensors move around the domain during training can be seen in Fig.~\ref{fig:seat_temp}. 
It is worth noting that the movement patterns of the sensors are not governed by physical principles, making their trajectories challenging to decipher. The sensors locations are tuned in conjunction with the network's trainable parameters, adding a layer of complexity to understanding their specific motion behavior.

\section{Discussion and conclusion}

We believe that the improved performance evidenced by our results may be attributed to a more effective utilization of training data, additional information gleaned from the interpolation process, and enhanced spatial awareness within the model gained during training.

This led us to a key realization: it was not merely the final positioning of the sensors that mattered, but the journey — or the dynamic positioning process — that truly played a pivotal role in model performance. 
This unexpected yet insightful outcome was further confirmed through additional experiments with the sea temperature dataset. Across multiple trainings using this dataset, the results consistently reinforced our revised understanding: optimal performance was achieved through dynamic, not static, sensor positioning. 


This approach, while superior in the task of field reconstruction, comes with an increased computational cost. However, it is important to clarify that this rise in computational demand is not attributed to an increase in the number of trainable parameters. Instead, it stems from the necessity to continuously compute the positional encoding and execute the interpolation step on the fly.
There are a number of potential improvements that can be readily identified for this workflow. On one hand, this framework makes use of linear interpolation, which may not perfectly capture the complexities of certain fields. This simplification, while beneficial for computational efficiency and ease of implementation, might introduce inaccuracies when dealing with high-frequency variations in the field.
Also, our study utilized sine-cosine encoding for sensor positions, which is not the only potential method of representation. Future research could explore the effectiveness of other position encodings, such as wavelet or even learned encodings, to determine if these can offer improved performance or additional benefits. 


In conclusion, this study introduces a fully end-to-end, differentiable framework capable of reconstructing complex fields from sparse representations. This framework has demonstrated an enhanced capability by allowing the adjustment of sparse sensor positions. Notably, the adaptive nature of our approach led to state-of-the-art results on a rigorous physics benchmark.
\section{Acknowledgments}
JES and AM gratefully acknowledge the support of the Center for Non-Linear Studies (CNLS) for this work.

\newpage
\appendix
\section{Appendix}

In this Appendix a few technical details regarding the developed optimization framework can be found.

\begin{figure}[H]
\centering
\includegraphics[width=1\textwidth]{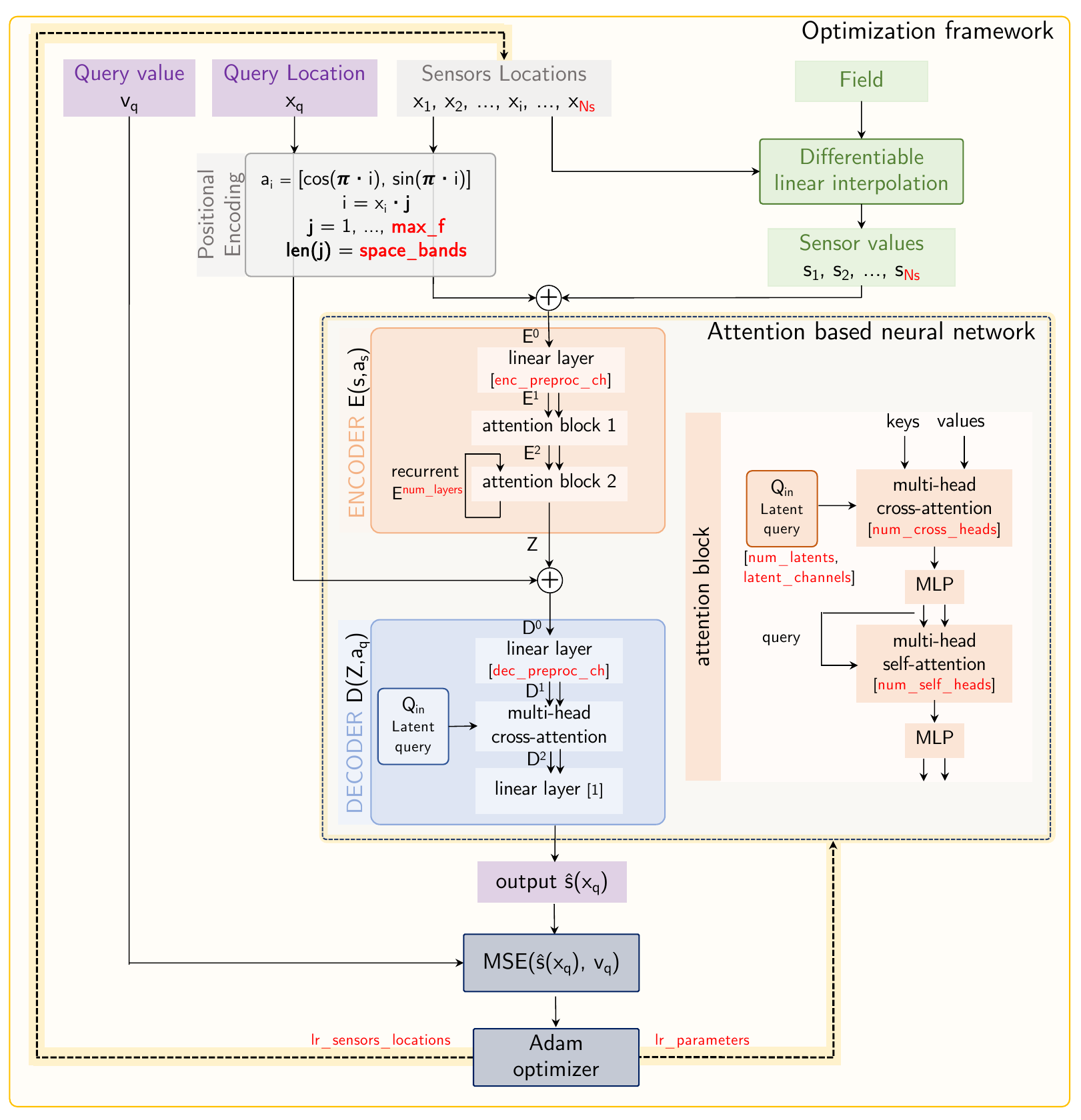}
    \caption{Optimization framework. The sensors position in the training set is a trainable parameter for the gradient descent optimizer, during each optimization step the parameters of the attention based neural network and the sensors positions are updated through this differentiable workflow. In red the hyperparameters, the values employed in this work are summarized in Tab. \ref{tab:hyperp}.}
    \label{fig:workflow_opt}
\end{figure}

In Fig.~\ref{fig:workflow_opt} the optimization framework is presented via a full schema of the data flow, expanding on the sketch present in Fig.~\ref{fig:workflow}, where also all the necessary hyperparameters are clearly indicated.
These, and their values chosen for this work, are summarized in Tab.~\ref{tab:hyperp}.

\textcolor{white}{asd}

\vspace{1cm}

\begin{table}[H]
\caption{Architecture and training hyperparameters.}
\label{tab:hyperp}
\vskip 0.15in
\begin{center}
\begin{small}
\begin{tabular}{cc}
\toprule
Hyperparameter   & Value  \\
\midrule
Ns     & 4-8-16 \\
max\_f     & 100 \\
space\_bands     & 32 \\
enc\_preproc\_ch     & 16 \\
num\_layers     & 3 \\
num\_cross\_head     & 16 \\
num\_latents & 256 \\
latent\_channels & 16 \\
num\_self\_head & 3 \\
dec\_preproc\_ch & 256 \\
$\mathrm{lr}_{\mathrm{sensors\_locations}}$ & 0.3 \\
$\mathrm{lr}_{\mathrm{parameters}}$ & 0.001 \\

\bottomrule
\end{tabular}
\end{small}
\end{center}
\vskip -0.1in
\end{table}

\bibliographystyle{plainnat.bst}
\bibliography{reference.bib}
\end{document}